\documentclass[aps, pre,amsmath,amssymb,twocolumn,showpacs,floatfix]{revtex4}

\usepackage{graphicx}
\usepackage{epsfig}

\newcommand {\dr}{{\mathrm d}\mathbf{r}}

\newcommand {\rr}{\mathbf{r}}

\begin{document}

\title{\bf Two-dimensional fluid with competing interactions exhibiting microphase separation: theory for bulk and interfacial properties}

\author{Andrew J.~Archer}
\affiliation{Department of Mathematical Sciences, Loughborough University, Loughborough, Leicestershire, LE11 3TU, United Kingdom}

\date{\today}

\begin{abstract} 

Colloidal particles that are confined to an interface such as the air-water interface are an example of a two-dimensional fluid. Such dispersions have been observed to spontaneously form cluster and stripe morphologies in certain systems with isotropic pair potentials between the particles, due to the fact that the pair interaction between the colloids has competing attraction {\em and} repulsion over different length scales. Here we present a simple density functional theory for a model of such a two-dimensional fluid. The theory predicts a bulk phase diagram exhibiting cluster, stripe and bubble modulated phases, in addition to homogeneous fluid phases. Comparing with simulation results for this model from the literature, we find that the theory is qualitatively reliable. The model allows for a detailed investigation of the structure of the fluid and we are able to obtain simple approximate expressions for the static structure factor and for the length scale characterising the modulations in the microphase separated phases. We also investigate the behaviour of the system under confinement between two parallel hard walls. We find that the confined fluid phase behaviour can be rather complex.

\end{abstract} 
\pacs{82.70.-y, 05.20.Jj, 61.20.Gy, 64.70.pv} 
\maketitle 
%\epsfclipon  

%\begin{multicols}{2}

\section{Introduction}

Many soft matter systems are capable of displaying spatially modulated structures. Well known systems such as solutions of amphiphiles and diblock copolymers can exhibit several microphase ordered phases \cite{SeulAndelmanScience1995}. This self organisation originates from competing interactions occurring over different length scales and particle orientations. Systems of colloidal particles confined to the air-water interface are also capable of exhibiting microphase separation \cite{BresmeOettelJPCM2007, GhezziEarnshawJPCM1997, SearetalPRE1999}. In these two dimensional fluids, the self organisation is driven by a competition between short ranged attractive interactions, which gather the particles together, but at longer ranges there is a repulsive interaction between the particles which breaks the system up into high density islands of particles with low density regions in between the islands. The shape of these islands depends on the surface coverage (density). At lower densities the particles form an array of (roughly) circular clusters, but at higher densities, they form a series of elongated parallel stripes \cite{GhezziEarnshawJPCM1997, SearetalPRE1999}. This self organisation is rather striking, when bearing in mind that the interactions between the individual pairs of particles is only a function of the particle separation and not of any orientational degrees of freedom, which play an important role in the microphase ordering of, for example, amphiphilic systems.

Suspensions of colloidal particles (i.e.\ three dimensional fluids) with effective pair interactions between the particles of a similar form have also been synthesised \cite{StradneretalNature2004, SedgwicketalJPCM2004, Bartlett, Bartlett2}. These systems also display various microphase separated phases including cluster phases. An overview of the various studies of models for these three dimensional systems is outlined in Ref.\ \cite{ARCHER21}.

The focus of this paper is a model two-dimensional fluid, first proposed in Ref.\ \cite{SearetalPRE1999}, and studied in detail via Monte Carlo computer simulations in a number of publications by Imperio and Reatto \cite{ImperioReattoJPCM2004, ImperioReattoJCP2006, ImperioPiniReatto2006, ImperioReatto2007}. This system is composed of particles with a hard core of diameter $\sigma$, interacting via the following pair potential:
\begin{equation}
v(r) = 
\begin{cases}
\infty \hspace{13mm} r \leq \sigma \\
w(r) \hspace{9.5mm} r > \sigma,
\end{cases}
\label{eq:pair_pot}
\end{equation}
where
\begin{equation}
w(r)=-\frac{\epsilon_a \sigma^2}{R_a^2} \exp \left(-\frac{r}{R_a} \right)
+\frac{\epsilon_r \sigma^2}{R_r^2} \exp \left(-\frac{r}{R_r} \right).
\label{eq:w}
\end{equation}
The parameter subscripts $a$ and $r$ stand for ``attraction'' and ``repulsion'' respectively. In much of the work presented here we will follow Imperio and Reatto and select the following values for the pair potential parameters: $R_a=1 \sigma$, $R_r=2 \sigma$, and $\epsilon_a=\epsilon_r=\epsilon$. Thus $\epsilon$ will be the unit of energy in the system. These parameters are chosen so that the integral over $w(r)$ is zero, i.e.\ $\alpha\equiv - \int \dr w(r)=0$.

The studies in Refs.\ \cite{SearetalPRE1999, ImperioReattoJPCM2004, ImperioReattoJCP2006, ImperioPiniReatto2006, ImperioReatto2007} found a rich phase behaviour displayed by this model fluid, which mirrored the behaviour observed in the experiments \cite{GhezziEarnshawJPCM1997, SearetalPRE1999}. At high temperatures, the contribution from $w(r)$ to $v(r)$ is negligible and so the fluid properties are wholly determined by the hard-core part of the potential $v(r)$ and the system is effectively just a fluid of hard-disks. On lowering the temperature, however, one finds that on increasing the fluid density at fixed temperature, there is a sequence of phase transitions. At low densities, the particles are uniformly distributed across the surface; we denote this the vapour phase. As the density is increased we find a transition to the cluster phase, in which the particles gather to form circular islands. The number of particles in each island may vary from just a few to as many as a hundred or more, depending on the particular values of the pair potential parameters in Eq.\ \eqref{eq:w}. The clusters are equilibrium structures and the probability of a particle moving from one cluster to another is $>0$. There is no long range order between the clusters -- the system forms a fluid of clusters. However, as the density is further increased, the clusters `freeze' so that the clusters arrange themselves onto a triangular superlattice \cite{ImperioReattoJCP2006}. On further increasing the density of particles on the surface, there is a transition to a phase in which the particles form parallel stripes \cite{SearetalPRE1999, ImperioReattoJPCM2004, ImperioReattoJCP2006}. Further increasing the density, finds the system exhibiting a `bubble' phase, in which there are low density voids amongst the particles on the surface. At even higher densities the system forms a uniform liquid phase \cite{ImperioReattoCom}. It is worth emphasising again that for some temperatures, these non-uniform structures are equilibrium (ergodic) states, having a non-zero probability for a particle to move from one cluster or stripe to another. However, as the temperature is decreased, this transition probability decreases. At low temperatures, one finds the same sequence of phase transitions, but for these temperatures the particles can be frozen within the clusters or stripes. The possibility of a related glass transition should not be ruled out either \cite{ImperioReattoJCP2006}.

It should also be recalled that pattern formation in two dimensional model fluids composed of particles interacting via isotropic pair potentials is not restricted to generalisations of Eqs.\ \eqref{eq:pair_pot} and \eqref{eq:w} -- i.e.\ potentials with a short ranged attraction and a longer ranged repulsion. Striped patterns have also been observed in systems of particles interacting via a pair potential having a hard-core plus a longer ranged purely repulsive `shoulder' potential \cite{MalescioPellicaneNature2003} and also in these systems when an additional attractive contribution is added to the potential, beyond the repulsive shoulder -- i.e.\ potentials with a short ranged repulsion and a longer ranged attraction \cite{MalescioPellicanePRE2004}.

In this paper we develop and apply a simple density functional theory (DFT) for the model fluid defined by Eqs.\ \eqref{eq:pair_pot} and \eqref{eq:w}. In Sec.\ \ref{sec:DFT}, we describe the DFT. In Sec.\ \ref{sec:structure} we use the DFT we investigate the structure of the fluid and, in particular, we obtain a simple expression for the static structure factor and also for the characteristic length scale associated with the modulations in the non-uniform phases. In Sec.\ \ref{sec:bulk_phase_behaviour} we calculate the phase diagram predicted by the simple DFT and find that it exhibits a cluster, stripe and bubble phase, which is in qualitative agreement with the simulation results of Refs.\ \cite{ImperioReattoJPCM2004, ImperioReattoJCP2006, ImperioPiniReatto2006}. In Sec.\ \ref{sec:inhom} we use the DFT to study the phase behaviour of the fluid when confined between two parallel hard walls. Depending on the fluid density and temperature, we find that as the separation between the two walls is varied, the equilibrium density profile can vary significantly due to the need for the length scale of the density modulations in the fluid to be commensurate with the distance between the walls. Finally, in Sec.\ \ref{sec:conc} we discuss our results and draw some conclusions.

\section{A DFT for the system}
\label{sec:DFT}

The structural and thermodynamic properties of the system may be obtained from the grand potential functional \cite{HM, Bob}:
\begin{equation}
\Omega[\rho(\rr)]={\cal F}[\rho(\rr)]+\int \dr \rho(\rr) (V_{ext}(\rr)-\mu),
\label{eq:grand_pot}
\end{equation}
 where $\rho(\rr)$ is the fluid one body density, $V_{ext}(\rr)$ is the external potential, $\mu$ is the chemical potential and $\cal{F}[\rho(\rr)]$ is the intrinsic Helmholtz free energy functional. The grand potential of the system for a given $\mu$ and $V_{ext}(\rr)$, denoted $\Omega$, is the minimal value of $\Omega[\rho(\rr)]$, and the equilibrium one-body density profile $\rho(\rr)$ is that which minimises the grand potential functional \cite{HM, Bob}. As usual, ${\cal F}[\rho(\rr)]$ is an unknown functional and we assume the following simple mean-field approximation for this quantity:
\begin{eqnarray}
{\cal F}[\rho(\rr)]&=&\int \dr \rho(\rr)f(\rho(\rr))\notag \\
&\,&+\frac{1}{2} \int \dr \int \dr' \rho(\rr) \rho(\rr') w(\rr-\rr')
\label{eq:F}
\end{eqnarray}
where $f(\rho)$ is the Helmholtz free energy per particle of a uniform fluid of hard-disks with bulk density $\rho$. This local density approximation (LDA) for the reference hard disk contribution includes the (exact) ideal gas contribution
\begin{equation}
{\cal F}_{id}[\rho(\rr)]=k_BT\int \dr \rho(\rr)(\ln[\Lambda^2 \rho(\rr)]-1),
\label{eq:F_id}
\end{equation}
where $\Lambda$ is the thermal De Broglie wavelength. For simplicity, we will use the scaled particle approximation,
\begin{equation}
\beta f(\rho)=\ln(\Lambda^2 \rho)-2-\ln(1-\eta)+(1-\eta)^{-1},
\label{eq:f_hd}
\end{equation}
where $\beta=1/k_BT$ is the inverse temperature and $\eta=\pi \rho \sigma^2/4$ is the packing fraction. There exist better approximations than Eq.\ (\ref{eq:f_hd}), such as that obtained in Ref.\ \cite{SantosetalJCP1995}, but for the densities of interest here, the more simple expression in Eq.\ (\ref{eq:f_hd}) is sufficiently reliable. The mean field contribution to the free energy in Eq.\ \eqref{eq:F}, from the tail of the pair potential, is justified on the basis that $w(r)$ is fairly long ranged and slowly varying. Note that Eq.\ (\ref{eq:F}) does not provide a reliable approximation for the Helmholtz free energy in cases when the fluid density $\rho(\rr)$ varies strongly on length scales $\sim \sigma$, i.e.\ for describing effects from packing of the hard cores of the particles. To take account of such correlations, one must implement a non-local approximation for the reference hard-disk free energy functional, along the lines of that used in Ref.\ \cite{ARCHER19} for the three-dimensional counterpart of the present system. However, in the present study, where our interest is in the cluster, stripe or bubble structures, for which the fluid density profile $\rho(\rr)$ varies over length scales $\gg \sigma$, Eq.\ (\ref{eq:F}) suffices \cite{Archerpreprint}.

\section{Structure of the uniform fluid}
\label{sec:structure}

The direct pair correlation function, defined as follows \cite{HM}:
\begin{equation}
c(\rr,\rr')=-\beta\frac{\delta^2({\cal F}[\rho(\rr)]-{\cal F}_{id}[\rho(\rr)])}{\delta \rho(\rr) \delta\rho(\rr')},
\label{eq:c_2}
\end{equation}
can be used to characterise two-point correlations in the inhomogeneous fluid. In the homogeneous bulk fluid, where $\rho(\rr)=\rho$, we find $c(\rr,\rr')=c(|\rr-\rr'|)=c(r)$. More conventionally one characterises the fluid structure by either considering the radial distribution function $g(r)$, obtained from  $c(r)$ via the Ornstein Zernike equation \cite{HM}, or the static structure factor
\begin{equation}
S(k)=\frac{1}{1-\rho \hat{c}(k)},
\label{eq:S_of_k}
\end{equation}
where $\hat{c}(k)$ is the Fourier transform of $c(r)$.

For the present model, Eqs.\ \eqref{eq:F} -- \eqref{eq:c_2} together yield the following expression for the pair direct correlation function
\begin{eqnarray}
c(\rr,\rr')&=&-\beta\left[2f'(\rho(\rr))+\rho(\rr)f''(\rho(\rr))-\frac{k_BT}{\rho(\rr)}\right] \delta(\rr-\rr')\notag \\
&\,&-\beta w(\rr-\rr'),
\label{eq:c_model}
\end{eqnarray}
where $\delta(\rr)$ is the Dirac delta function and $f'$ and $f''$ are the first and second derivatives of $f$ with respect to $\rho$. Combining Eqs.\ \eqref{eq:f_hd} -- \eqref{eq:c_model} we obtain the following expression for the static structure factor:
\begin{equation}
S(k)=\frac{1}{a(\eta)+\rho \beta \hat{w}(k)},
\label{eq:S_of_k_model}
\end{equation}
where $a(\eta)=(1+\eta)/(1-\eta)^3$ and where
\begin{equation}
\hat{w}(k)=-\frac{2 \pi \epsilon_a \sigma^2}{(1+R_a^2 k^2)^{3/2}}
+\frac{2 \pi \epsilon_r \sigma^2}{(1+R_r^2 k^2)^{3/2}}.
\label{eq:11}
\end{equation}
Due to the local density approximation for the hard-disk contribution to the free energy functional (\ref{eq:F}), this expression for $S(k)$ is strictly only applicable for wave numbers $k<2 \pi/\sigma$. In Ref.\ \cite{ImperioReattoJPCM2004}, the authors showed that the RPA approximation for $S(k)$ is fairly reliable for describing the fluid structure. The RPA consists of making the approximation $\hat{c}(k)=\hat{c}_{hd}(k)-\beta \hat{w}(k)$, where $\hat{c}_{hd}(k)$ is the Fourier transform of the hard disk pair direct correlation function, $c_{hd}(r)$. For the present model fluid, the RPA approximation yields a structure factor which is reliable for all wave numbers $k$. The result in Eq.\ \eqref{eq:S_of_k_model} only coincides with the RPA result for small wave numbers $k<2 \pi/\sigma$. Thus the present theory effectively only includes the inter-cluster correlation contributions to $S(k)$ and does not include the intra-cluster correlation contributions.

As an aside, we now briefly consider the structure and phase behaviour of a system for which $\epsilon_r=0$, i.e.\ when the long range repulsive contribution is no longer present and $(-\alpha)=\int \dr w(r)=\hat{w}(k=0)=-2 \pi \epsilon_a \sigma^2<0$. In this case, the system exhibits only bulk vapour-liquid phase coexistence and no microphase separation. Owing to our simple approximation for the free energy \eqref{eq:F}, it is straightforward to determine the phase diagram and calculate the binodal and spinodal. For example, the spinodal corresponds to the divergence $S(k=0) \rightarrow \infty$; i.e.\ this is when the denominator in Eq.\ (\ref{eq:S_of_k_model}) equals zero for $k=0$, giving the following expression for the spinodal temperature as a function of density:
\begin{equation}
\frac{k_BT}{\alpha}=\frac{\rho(1-\eta)^3}{(1+\eta)}.
\label{eq:spinodal}
\end{equation}
The critical point corresponds to the maximum of this curve, i.e.\ when $\frac{\partial}{\partial \rho } [ \rho(1-\eta)^3/(1+\eta)]=0$. This yields a critical packing fraction $\eta_c=(\sqrt{7}-2)/3 \simeq 0.274$. Note that one could also obtain Eq.\ (\ref{eq:spinodal}) starting directly from the free energy functional \eqref{eq:F}. In bulk, this gives the following expression for the bulk free energy per particle: $f_{tot}(\rho)=f(\rho)-\rho \alpha/2$. From this, we may obtain the pressure in the system:
\begin{eqnarray}
P&=& \rho^2\left(\frac{\partial f_{tot}}{\partial \rho} \right) \notag \\
&=&\frac{\rho k_BT}{(1-\eta)^2}-\frac{\rho^2 \alpha}{2}.
\label{eq:pressure}
\end{eqnarray}
The spinodal corresponds to the locus $(\partial P/\partial \rho)_T=0$, so taking a further derivative we obtain Eq.\ (\ref{eq:spinodal}). Such consistency between the structural and the thermodynamic routes to the phase behaviour comes as a consequence of deriving all quantities from a free energy functional and is one of the advantages of using the DFT approach rather than integral-equation theory based approaches \cite{ARCHER21, HM, Caccamo}.

We return now to the more general case when $\epsilon_r \neq 0$. First, we focus on the small wave number $k$ behaviour of $S(k)$. Taylor expanding $S(k)$ in Eq.\ (\ref{eq:S_of_k_model}) around $k=0$ we find:
\begin{equation}
S(k)=S_0-\gamma S_0^2 k^2+{\cal O}(k^4) 
\label{eq:S_of_k_model_expand}
\end{equation}
where $S_0=S(k=0)=[a(\eta)-\rho \beta \alpha]^{-1}$ and $\gamma=3 \pi \sigma^2 \rho \beta (\epsilon_a R_a^2- \epsilon_r R_r^2)$. When $\gamma>0$, i.e.\ when $(\epsilon_a R_a^2- \epsilon_r R_r^2)>0$, then $S(k=0)$ is a maximum and the phase behaviour of the system is the same as in the case when $\epsilon_r=0$, provided $\alpha>0$, where $\alpha=-\int \dr w(r)=2 \pi \sigma^2 (\epsilon_a-\epsilon_r)$, and the spinodal curve is given by Eq.\ (\ref{eq:spinodal}). However, when $\gamma<0$, i.e.\ when $(\epsilon_a R_a^2- \epsilon_r R_r^2)<0$, then $S(k=0)$ is a minimum and there is a maximum in $S(k)$ at $k=k_c \neq 0$. This peak in $S(k)$ at $k_c$ indicates a propensity in the fluid for having density modulations with wavelength $2\pi/k_c$. In fact, in certain portions of the phase diagram, the fluid may become unstable with respect to density fluctuations of wavelength $2 \pi/k_c$ and it then follows that the system exhibits microphase separation to the various different modulated fluid phases, in this portion of the phase diagram. This instability is indicated by the divergence of the peak in $S(k)$ at $k_c$. We denote the locus in the phase diagram at which this occurs the $\lambda$-line -- i.e.\ the $\lambda$-line is defined as the locus where $S(k=k_c)\rightarrow \infty$ \cite{ARCHER19, Archerpreprint, StellJSP1995, CiachetalJCP2003, PatsahanCiachJPCM2007, Archer6}.

Within the present DFT we may obtain an expression for $k_c$ and also for the temperature on the $\lambda$-line, as a function of the density, as follows: the wave number $k_c$ is defined as the value of $k$ where $S(k)$ is a maximum, i.e.\ where the quantity $[a(\eta)+\rho\beta\hat{w}(k)]$ is a minimum [recall Eq.\ (\ref{eq:S_of_k_model})]. This occurs when $\frac{\partial}{\partial k}[a(\eta)+\rho\beta\hat{w}(k)]=\rho\beta\frac{\partial}{\partial k}\hat{w}(k)=0$. Solving this equation we find that there are two solutions. The first at $k=0$, corresponding to the minimum in $S(k)$ and the second at $k=k_c$, where
\begin{equation}
k_c=\sqrt{\frac{\Gamma-1}{R_r^2-\Gamma R_a^2}},
\label{eq:k_c}
\end{equation}
and where $\Gamma=(\epsilon_r R_r^2/\epsilon_a R_a^2)^{2/5}$. For the set of pair potential parameters that we focus on here, $R_a=1 \sigma$, $R_r=2 \sigma$, and $\epsilon_a=\epsilon_r=\epsilon$, we find $k_c \sigma \simeq 0.573$ so that the length-scale characterising the density modulations in the microphases $=2 \pi/k_c \simeq 11\sigma$, as pointed out in Ref.\ \cite{ImperioReattoJPCM2004}. Comparing the result in Eq.\ \eqref{eq:k_c}, with the results from Monte-Carlo computer simulations, for various different sets of choices for the pair potential parameters, one finds that Eq.\ \eqref{eq:k_c} is generally reliable and in good agreement with the simulation results \cite{ImperioReattoCom}.

The $\lambda$-line corresponds to the locus where $[a(\eta)+\rho\beta\hat{w}(k=k_c)]=0$. Rearranging this we obtain the following expression for the temperature along the $\lambda$-line:
\begin{equation}
\frac{k_BT}{\Delta}=\frac{\rho(1-\eta)^3}{(1+\eta)},
\label{eq:lambda_line}
\end{equation}
where $\Delta=-\hat{w}(k=k_c)>0$. Note the strong similarity between this expression and Eq.\ (\ref{eq:spinodal}) for the spinodal. For the pair potential parameters $R_a=1 \sigma$, $R_r=2 \sigma$, and $\epsilon_a=\epsilon_r=\epsilon$, the $\lambda$-line is displayed in Fig.\ \ref{fig:phase_diag}. See also Fig.\ 14 in Ref.\ \cite{ImperioReattoJPCM2004}, where the $\lambda$-line from the full RPA is displayed. Inside the $\lambda$-line, the homogeneous fluid is unstable with respect to density fluctuations with wave number $k_c$. The nature of the phases that occur in this region of the phase diagram is the topic of the following section.

\section{Bulk phase behaviour}
\label{sec:bulk_phase_behaviour}

\begin{figure}
\includegraphics[width=1.\columnwidth]{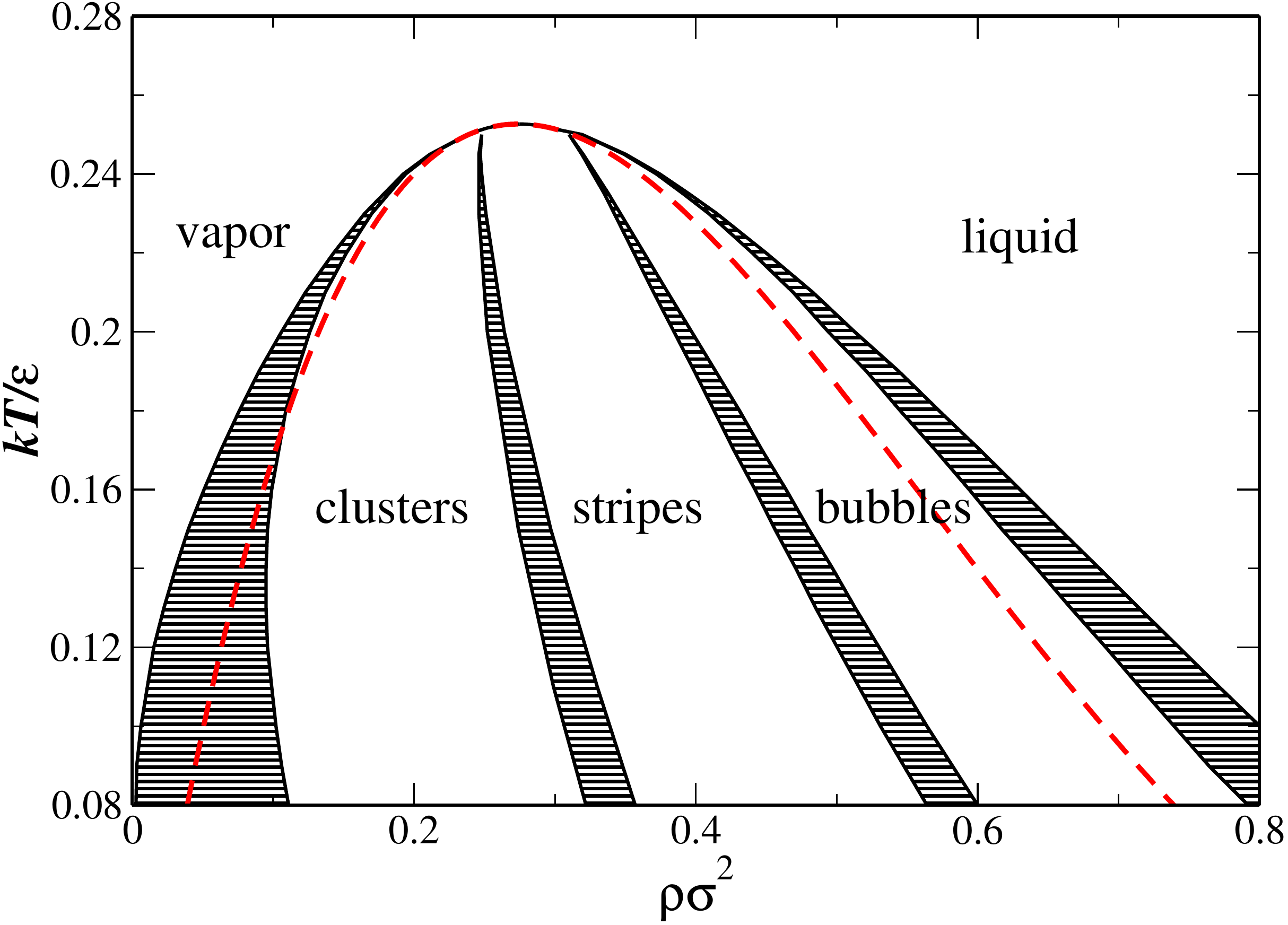}
\caption{\label{fig:phase_diag}
(Color online) Phase diagram in the density-temperature plane. The shaded regions denote two-phase coexistence regions between the various phases and the dashed line is the $\lambda$-line.}
\end{figure}

\begin{figure}
\includegraphics[width=7cm,height=5cm]{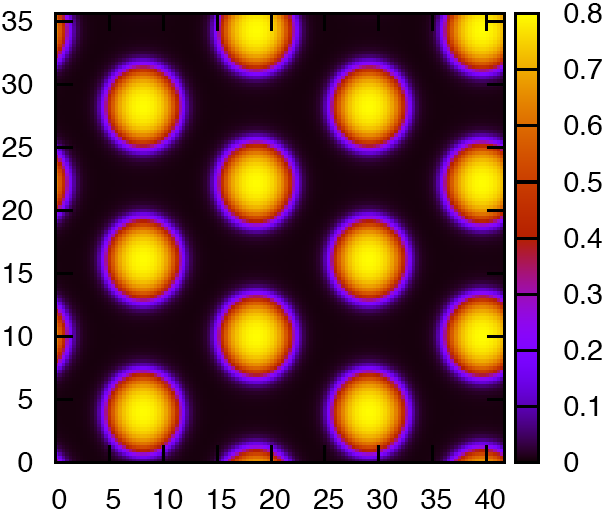}

\includegraphics[width=7cm,height=5cm]{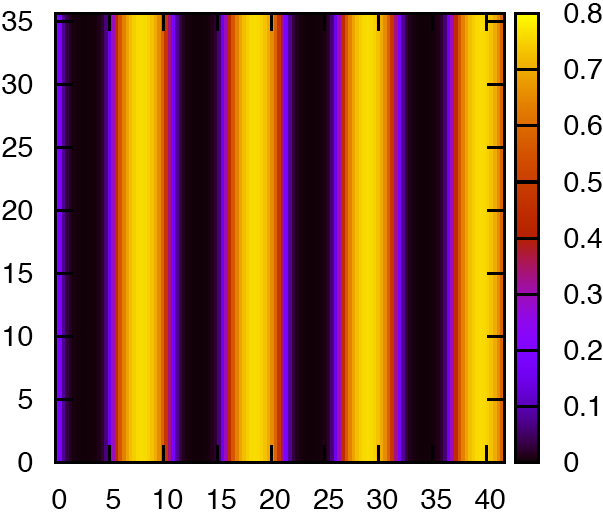}

\includegraphics[width=7cm,height=5cm]{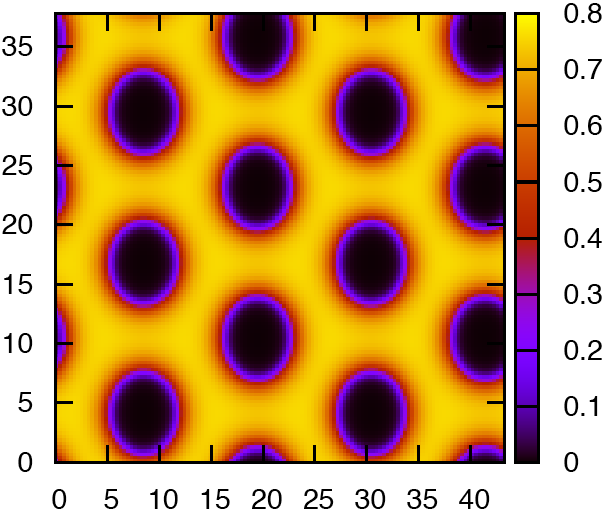}
\caption{\label{fig:density_profiles}
(Color online) Typical density profiles in each of the microphase separated phases. These profiles are for the temperature $k_BT/\epsilon=0.15$ and the average densities are $\rho \sigma^2=0.2$, 0.35 and $0.5$ in the cluster, stripe and bubble phases respectively. The $x$- and $y$-axes are in units of $\sigma$.}
\end{figure}

For a given state point, the equilibrium fluid density profile is that which minimises the grand potential (\ref{eq:grand_pot}); i.e.\ it is the solution to the following equation \cite{HM, Bob}:
\begin{equation}
\frac{\delta \Omega[\rho(\rr)]}{\delta \rho(\rr)}=0.
\label{eq:EL_eq}
\end{equation}
From Eqs.\ \eqref{eq:grand_pot}, \eqref{eq:F} and \eqref{eq:EL_eq} we obtain
\begin{equation}
f(\rho(\rr))+\rho(\rr)f'(\rho(\rr))+\int \dr' \rho(\rr') w(\rr-\rr')+V_{ext}(\rr)=\mu.
\label{eq:EL_eq_model}
\end{equation}
We discretise the density profile $\rho(\rr)$ on a two-dimensional Cartesian grid and, for a given value of the chemical potential $\mu$, we may then solve Eq.\ \eqref{eq:EL_eq_model} for the fluid density profile, by using a simple iterative numerical procedure. For the bulk case, where the external potential $V_{ext}(\rr)=0$, we solve using periodic boundary conditions in both Cartesian directions. Since in this case the uniform density profile $\rho(\rr)=\rho$ is always a solution of Eq.\ \eqref{eq:EL_eq_model}, to find the density profile for the modulated phases we investigated three approaches: (i) choose an initial guess for the input density profile with a cluster/stripe/bubble already present, to `nucleate' other clusters/stripes/bubbles throughout the remainder of the system. (ii) Set the initial input density profile to be
\begin{equation}
\rho(\rr)=\rho+\chi(\rr),
\label{eq:rho_noise}
\end{equation}
where $\chi(\rr)$ is a random noise field with the property $|\chi(\rr)| \ll \rho$. This is the ``quick and dirty" way of establishing the fluid phase behaviour for a particular $\rho$. (iii) Take the density profile obtained previously for a neighbouring state point and use it as an initial guess for the density profile at the required state point. Approaches (i) and (ii) proved to be good for an initial assessment of the phase behaviour. However, none of these approaches guarantees that one finds the density profile corresponding to the global minimum of the grand potential. To find the true global minimum one must follow approach (iii) above for all the possible candidate structures and then take the resulting density profiles (which each correspond to local minima in the free energy landscape) and substitute the density profiles into the grand potential functional (\ref{eq:grand_pot}) and evaluate the grand potential $\Omega$ corresponding to each phase. The phase with the lowest grand potential is the true equilibrium phase. Note also that since we were solving for the density profile on a finite grid of size $L_x \times L_y$, with periodic boundary conditions, one must also minimise the grand potential with respect to $L_x$ and $L_y$ as well as with respect to variations in $\rho(\rr)$, since otherwise one is forcing the period of the fluid modulations to be $L_x/n$ in the $x$-direction and $L_y/m$ in the $y$-direction, where $n$ and $m$ are integers \cite{footnote}. This process is not as arduous as it might first appear. This is due to the fact that, as noted in the previous section, the length scale determining the size of the periodic structures is $2 \pi/k_c$, which for the present DFT does not vary as a function of density or temperature. This means that having found the values of $L_x$ and $L_y$ which minimise $\Omega$ for a certain state point, the values of $L_x$ and $L_y$ which minimise $\Omega$ for a different point in the phase diagram are very close to those at the first state point.

In order to locate the phase transitions between the various phases and calculate the densities of the coexisting phases, we performed scans of the chemical potential $\mu$ for fixed temperature $T$. Along these scans we recorded the pressure $P$ (obtained from the grand potential, since $\Omega=-PA$, where $A$ is the system area) and the average density of the system $\rho$. We repeat this for all the phases displayed by the system. Since phase coexistence occurs between points with equal $(T,P,\mu)$ one can then read off the coexisting state points by plotting $P$ versus $\mu$ for these isothermal scans and noting the intersection points between the different curves. The resulting phase diagram is displayed in Fig.\ \ref{fig:phase_diag}. In addition to the homogeneous vapour and liquid phases we find that the DFT predicts that the system displays three inhomogeneous fluid phases. Examples of density profiles obtained for each of these three phases are displayed in Fig.\ \ref{fig:density_profiles}. At lower densities, the system forms a regular array of clusters. The density within the clusters is close to that of the uniform liquid for the same temperature and the density in the voids is low -- close to that of the uniform vapour for the same temperature. At intermediate densities, the system forms an array of parallel stripes. Again, the particle density within the stripes is that of a liquid and the local density between the stripes is that of the vapour. At higher densities still, the system forms a regular array of bubbles, i.e.\ within the bubbles the density of particles is low, but in between the bubbles the local density is that of the liquid.

All of the phase transitions are predicted by the present DFT to be first order, except for the transition from the uniform fluid to the stripe phase. This phase transition is predicted to be second order (continuous). If one starts in the uniform fluid phase at a state point directly above the stripe phase in temperature, and then decreases the temperature, one finds that the system is uniform right up to the $\lambda$-line, which is itself the phase transition line. The relevant order parameter for the phase transition is the amplitude of the stripe modulations ${\cal A}$ \cite{Archerpreprint}. Outside the $\lambda$-line, and on the $\lambda$-line itself, the amplitude ${\cal A}=0$. On decreasing the temperature below $T_\lambda$, the temperature on the $\lambda$-line, one finds that the amplitude of the modulations in the stripe phase grows continuously with decreasing temperature \cite{Archerpreprint}. We will discuss this issue in further detail in Sec.\ \ref{sec:conc}.

\section{Confined fluid properties}
\label{sec:inhom}

In this section we apply the DFT \eqref{eq:F} to investigate the behaviour of the fluid under confinement. In particular, we examine the case when the fluid is confined between two parallel hard walls, where the external potential varies in only one Cartesian direction and $V_{ext}(\rr)=V_{ext}(x)$, where
\begin{equation}
V_{ext}(x) = 
\begin{cases}
\infty \hspace{10mm} x \leq 0 \\
0 \hspace{9mm} 0<x <L \\
\infty \hspace{10mm} x \geq L
\end{cases}
\label{eq:ext_pot}
\end{equation}
This situation was studied in Ref.\ \cite{ImperioReatto2007} by Imperio and Reatto using Monte Carlo computer simulations. As in the previous section, we calculate the equilibrium fluid density profile $\rho(\rr)$ by discretising it on a two-dimensional Cartesian grid and then solving Eq.\ \eqref{eq:EL_eq_model} via a simple iterative numerical procedure, starting from an initial guess for the density profile. This is done using periodic boundary conditions in the Cartesian $y$-direction (the direction parallel to the walls).

For a given temperature $T$, we investigate the various morphologies that the fluid density profile displays as the distance between the two walls, $L$, is increased. These calculations could be performed for fixed chemical potential, $\mu$. However, the Monte-Carlo simulation results in Ref.\ \cite{ImperioReatto2007} were performed in the $NVT$ ensemble (fixed number of particles $N$, volume $V$ and temperature $T$), keeping the average particle density between the walls fixed, as $L$ was varied. In order to compare with these results, we perform our calculations at fixed average density $\rho$. Thus, for each value of $L$, the chemical potential is chosen so as to achieve this target surface density. This is done by renormalising the density profile to the desired value at each step in the iterative routine for calculating the density profile. This method means that one does not have to know {\it a priori} the precise value of $\mu$ required to achieve the desired average density.

\begin{figure}
\includegraphics[width=1.\columnwidth]{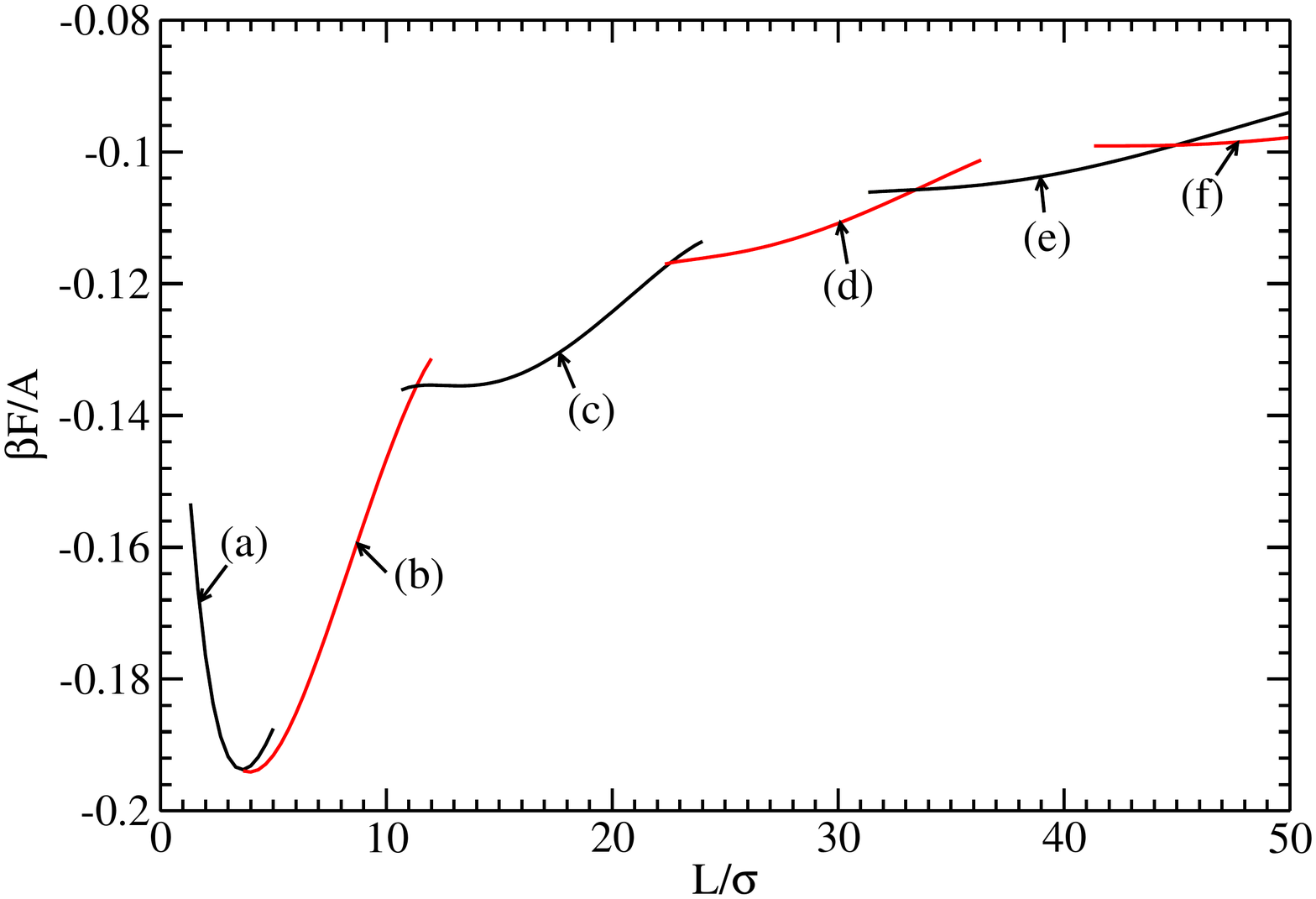}
\includegraphics[width=1.\columnwidth]{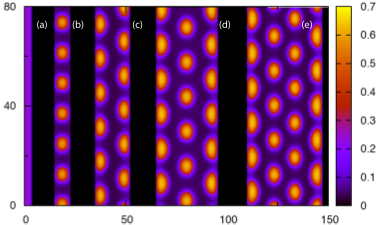}
\caption{\label{fig:slit_0_2}
(Color online) Top: the Helmholtz free energy per unit area, ${\cal F}/A$, as a function of slit width, $L$, for the case when the average fluid density in the slit is $\rho \sigma^2=0.2$ and the temperature $k_BT/\epsilon=0.2$. Bottom: a series of density profiles for increasing $L$, labelled (a) -- (e) which correspond to the different free energy curves in the upper figure, which are correspondingly labelled.}
\end{figure}

\begin{figure}
\includegraphics[width=1.\columnwidth]{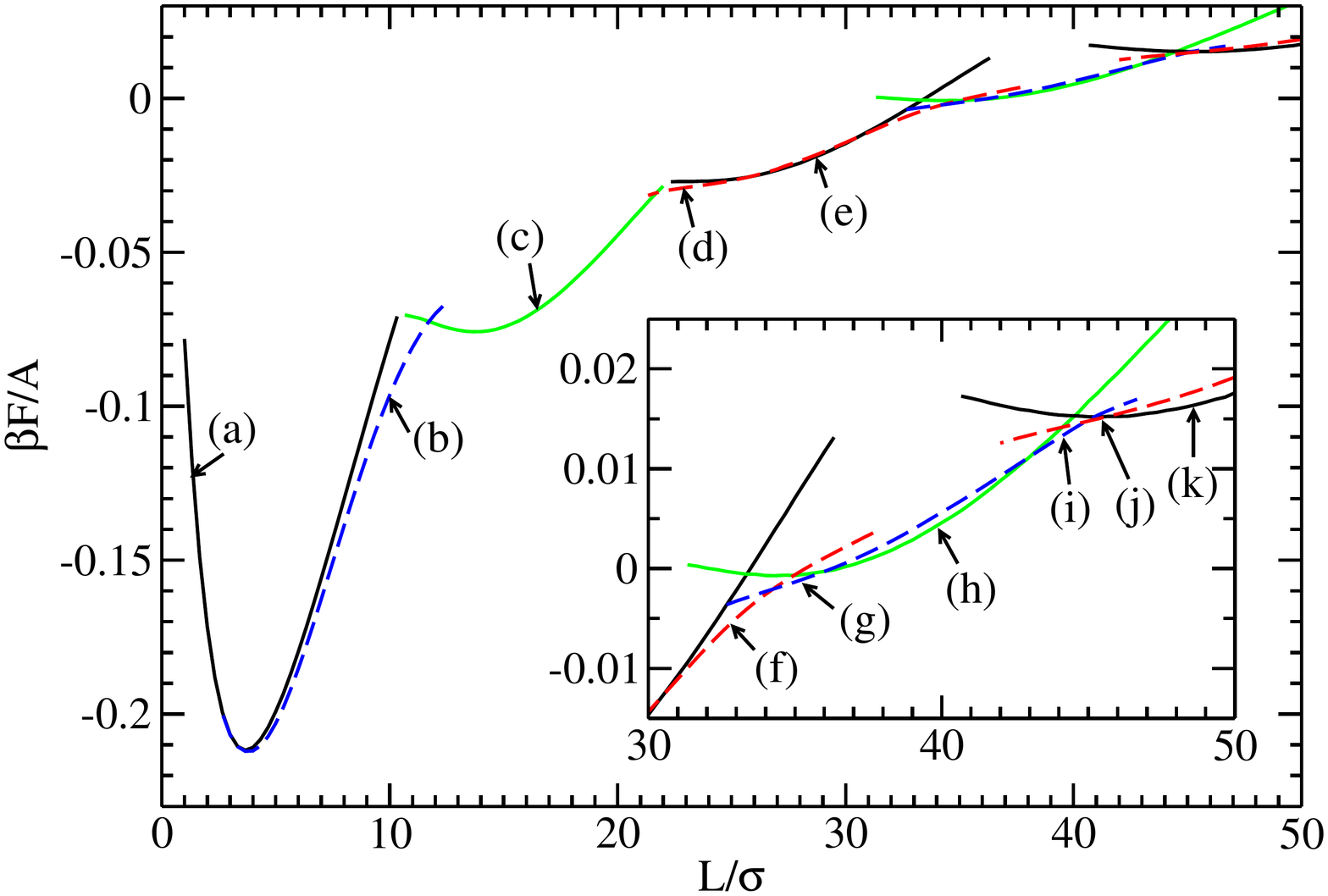}
\includegraphics[width=1.\columnwidth]{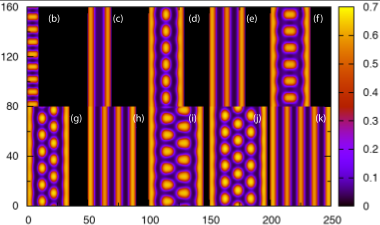}
\caption{\label{fig:slit_0_3}
(Color online) Same as Fig.\ \ref{fig:slit_0_2}, except here the average density in the slit is $\rho \sigma^2=0.3$.}
\end{figure}

\begin{figure}
\includegraphics[width=8cm,height=6.8cm]{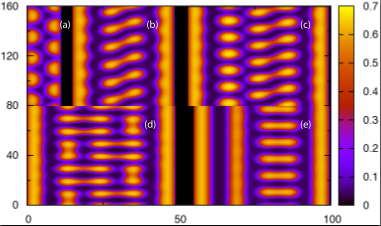}
\caption{\label{fig:metastable}
(Color online) Some of the metastable density profiles that were found for the case when $k_BT/\epsilon=0.2$ and the average density $\rho \sigma^2=0.3$.}
\end{figure}

For a given $L$, one finds that there can be several different density profiles which correspond to minima of the free energy -- such as, for example, one density profile exhibiting stripes parallel to the walls and another with stripes lying perpendicular to the walls. As in the previous section, in order to find the various candidate density profiles we implemented the following three different approaches: (i) choose an initial guess for the input density profile with a cluster/stripe/bubble already present, to nucleate other clusters/stripes/bubbles throughout the remainder of the system. (ii) Starting from the noisy initial density profile in Eq.\ \eqref{eq:rho_noise}. (iii) Take the density profile obtained previously for a neighbouring state point or value of $L$ as the initial guess. The approach taken was to use methods (i) and (ii) to generate the set of candidate structures and then we used method (iii), to calculate the Helmholtz free energy via Eq.\ \eqref{eq:F}, as a function of $L$.

In Fig.\ \ref{fig:slit_0_2}, we display the Helmholtz free energy per unit of available area between the walls $A$, as a function of the distance between the walls $L$ (upper figure), together with a sequence of representative density profiles (lower figure). The temperature is $k_BT/\epsilon=0.2$ and the average density is $\rho \sigma^2=0.2$, which as can be seen from Fig.\ \ref{fig:phase_diag} corresponds in the bulk to the cluster phase. For small values of $L < 2 \pi/k_c$, the fluid density profile does not vary in the $y$-direction (parallel to the walls). The branch of the free energy corresponding to this case is labelled (a) in the upper part of Fig.\ \ref{fig:slit_0_2} and a typical density profile for this branch of the free energy is correspondingly labelled in the lower part of Fig.\ \ref{fig:slit_0_2}. As $L$ is increased, we find a new minimum of the free energy, which corresponds to a density profile exhibiting a single line of clusters -- see the density profile labelled (b) in the lower part of Fig.\ \ref{fig:slit_0_2} and the correspondingly labelled free energy curve above. On further increasing $L$, one finds a sequence different free energy branches, with the density profile along each or these, having an additional line of clusters between the walls. Free energy curve (c) corresponds to two lines of clusters lying parallel to the walls, (d) to three lines of clusters, (e) to four lines of clusters and (f) to five lines of clusters (the corresponding density profile is not displayed in this case).

In Fig.\ \ref{fig:slit_0_3}, we display the Helmholtz free energy per unit area and a number of typical density profiles for the case when $k_BT/\epsilon=0.2$ and $\rho \sigma^2=0.3$. In bulk, this state point lies within the stripe phase and so for most values of $L$ the equilibrium density profile consists of stripes that lie parallel to the walls. However, for values of $L$ where the free energies for a density profile with $n$ and for a density profile with $(n+1)$ stripes are equal (i.e.\ when $L$ is incommensurate with the stripes), then one observes that the density profiles exhibiting clusters have a lower free energy and are the equilibrium configurations. Thus, even though for this average density the bulk phase is the stripe phase, due to the fact that this state point is not too far in the phase diagram from the cluster phase, one sees the influence of this proximity in the confined fluid density profiles. For small values of $L$, the fluid density profiles do not vary in the direction parallel to the walls -- the branch of the free energy corresponding to this is labelled (a) in the upper part of Fig.\ \ref{fig:slit_0_3}. On increasing $L$ we find (b), corresponding to short stripes lying perpendicular to the wall, followed by (c) where we find two stripes lying parallel to the walls. Increasing $L$, we find (d) corresponding to two parallel stripes with a line of clusters in between. Further increasing $L$ we find (e), corresponding to three stripes lying parallel to the walls. As $L$ is increased even further we see an increasingly complex series of transitions between configurations with different morphologies -- see the magnification in the inset of the upper panel of Fig.\ \ref{fig:slit_0_3}. Note that the density profiles displayed in Fig.\ \ref{fig:slit_0_3} are the ones that correspond to equilibrium density profiles (i.e.\ to global minima of the free energy, for some value of $L$). However, we also find a large number of other density profiles with higher free energies; see for example some of these displayed in Fig.\ \ref{fig:metastable}. These metastable configurations correspond to local minima in the free energy that are not global minima.

\begin{figure}
\includegraphics[width=1.\columnwidth]{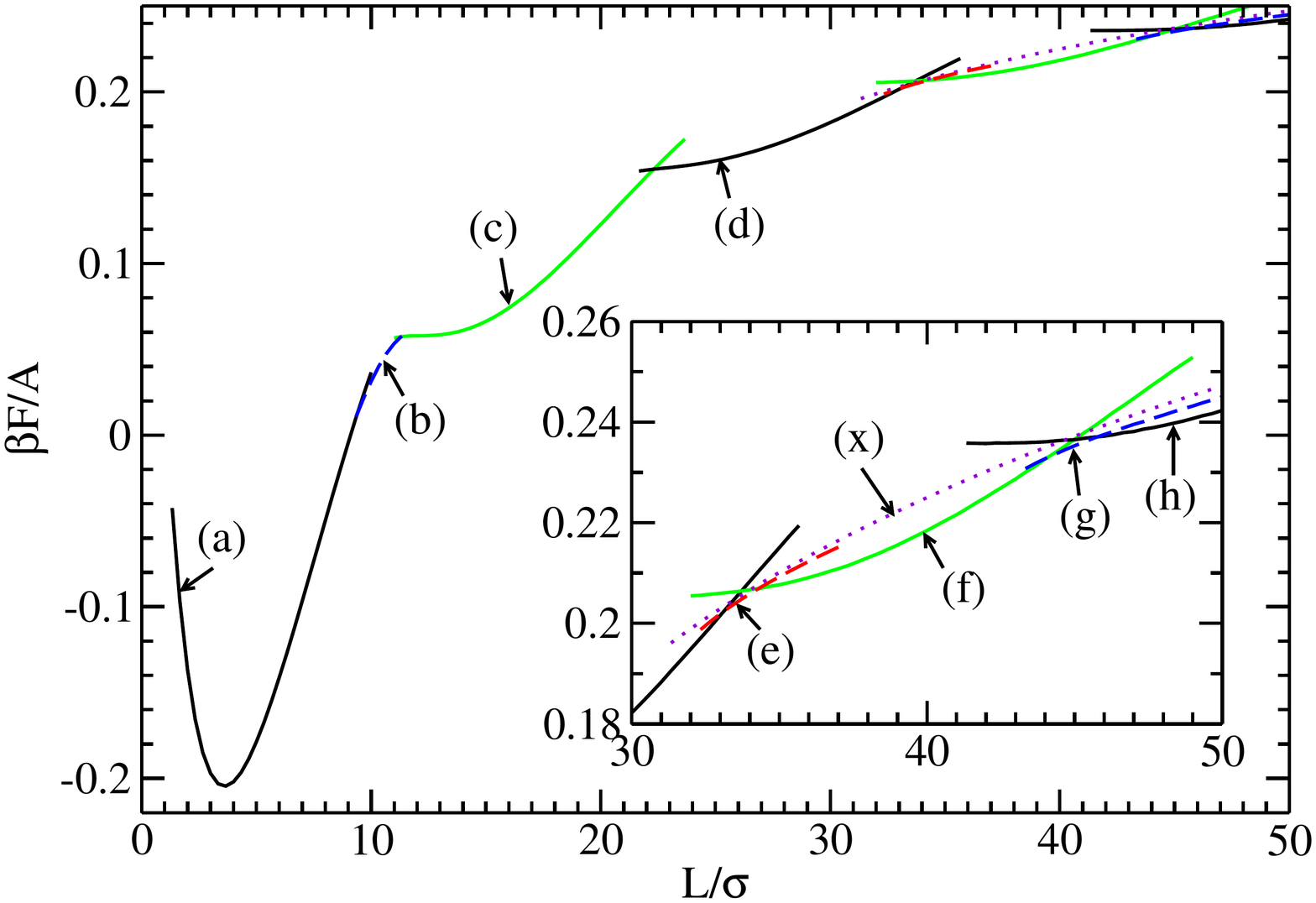}
\includegraphics[width=1.\columnwidth]{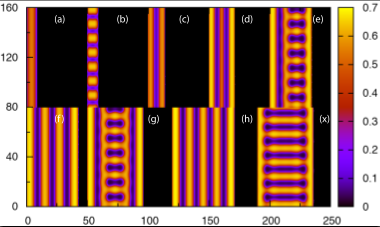}
\caption{\label{fig:slit_0_4}
(Color online) Same as Fig.\ \ref{fig:slit_0_2}, except here the average density in the slit is $\rho \sigma^2=0.4$.}
\end{figure}

\begin{figure}
\includegraphics[width=1.\columnwidth]{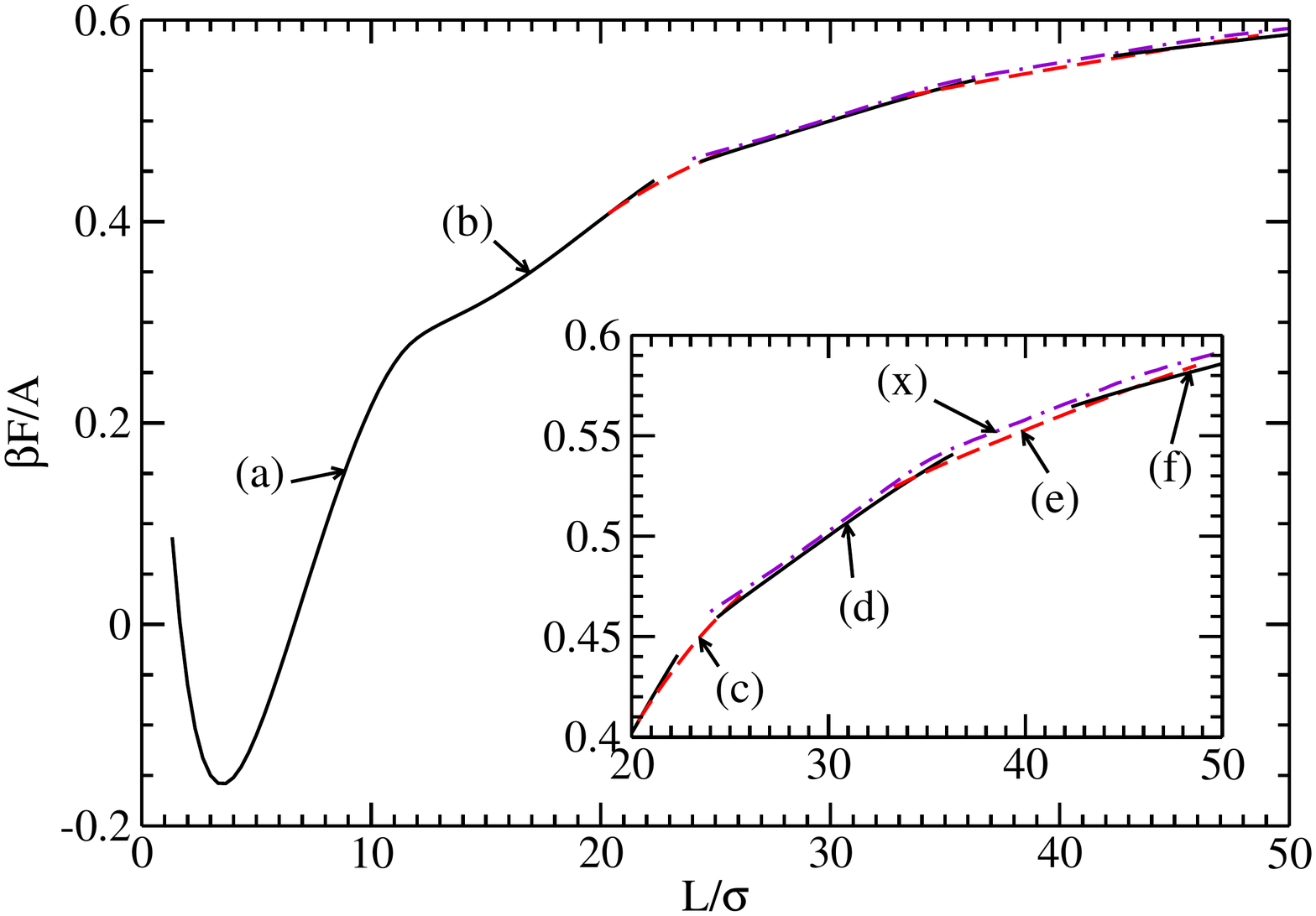}
\includegraphics[width=1.\columnwidth]{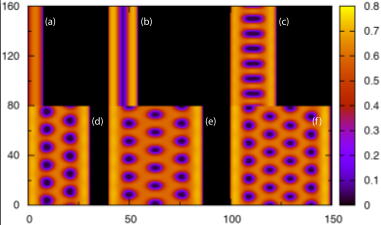}
\caption{\label{fig:slit_0_5}
(Color online) Same as Fig.\ \ref{fig:slit_0_2}, except here the average density in the slit is $\rho \sigma^2=0.5$.}
\end{figure}

In Fig.\ \ref{fig:slit_0_4}, we display the free energy per unit area together with a number of representative density profiles for the case when $k_BT/\epsilon=0.2$ and $\rho \sigma^2=0.4$. In bulk, this corresponds to a state point in the bubble phase, lying very close to the transition to the stripe phase. However, due to the influence of the confining walls, we find that for most values of $L$ investigated here, the equilibrium density profiles consist of stripes lying parallel to the walls. Thus, the confining walls exert a strong influence on the symmetry of the fluid density profiles, for this state point. Note that the `ladder' density profile labelled (x) corresponds to a configuration that does not correspond to the global minimum to the free energy. It is included in Fig.\ \ref{fig:slit_0_4} because for some values of $L$ it has a free energy that is very close to that of the true minima profiles, (e) and (g) -- see the magnification in the inset of the upper panel of Fig.\ \ref{fig:slit_0_4}.

In Fig.\ \ref{fig:slit_0_5}, we display results for the case when $k_BT/\epsilon=0.2$ and $\rho \sigma^2=0.5$. In bulk, this corresponds to a state point in the bubble phase, lying well removed from the transition to the stripe phase. However, due to the influence of the confining walls, for small values of $L$ we find that the equilibrium density profiles consist of stripes lying parallel to the walls. Note also that the transition from (a), having one stripe, to (b), having two stripes, appears to be a continuous transition.  The metastable free energy branch labelled (x) in Fig.\ \ref{fig:slit_0_5} is that corresponding to three, four and then five (as $L$ is increased) stripes lying parallel to the walls. The transition from three to four stripes and also from four to five stripes appear to be continuous transitions. Interestingly, this is not the case for the transition between two and three stripes. For $L/\sigma \gtrsim 21$ the equilibrium density profiles contain lines of bubbles lying parallel to the walls.

In Ref.\  \cite{ImperioReatto2007} Imperio and Reatto used Monte-Carlo computer simulations to investigate the behaviour of the present model fluid under confinement. Due to the fact that there is only qualitative agreement between the bulk phase diagram predicted by the present DFT and that obtained from simulations, it is hard to compare results directly. In Fig.\ 1 of their paper, Imperio and Reatto display the potential energy as a function of $L$ and also display a sequence of simulation snapshots for the case when $\rho\sigma^2=0.4$ and the temperature $T \simeq 0.8T_{max}$, where $T_{max}\simeq 0.13\epsilon/k_B$ is the maximum temperature at which modulated phases are observed to occur in the simulation results (for the present DFT, $T_{max}\simeq0.26\epsilon/k_B$; see Fig. \ref{fig:phase_diag}). The sequence of structures displayed in Fig.\ 1 of Ref.\ \cite{ImperioReatto2007} do not correspond exactly with any of the results displayed here in Figs.\ \ref{fig:slit_0_2}, \ref{fig:slit_0_3}, \ref{fig:slit_0_4} or \ref{fig:slit_0_5}. However, for values of $L< 30 \sigma$, the sequence of phases observed in the simulation results displayed in Fig.\ 1 of Ref.\ \cite{ImperioReatto2007} are the same as those displayed in Fig.\ \ref{fig:slit_0_3}. From this comparison, and also from comparing with the results in Fig.\ \ref{fig:slit_0_4}, which is a state point with the same average density as the results in Fig.\ 1 of Ref.\ \cite{ImperioReatto2007}, we conclude that the results in Ref.\ \cite{ImperioReatto2007} effectively correspond to a state point somewhere in-between the two state points for which results are displayed in Figs.\ \ref{fig:slit_0_3} and \ref{fig:slit_0_4}. Since a detailed comparison between our results and the simulation results of Ref.\ \cite{ImperioReatto2007} is not possible, we are only able to draw the general conclusion that the present theory appears to be qualitatively reliable for describing the confined fluid properties of the present system and seems to at least be able to describe some of the sequences of structures that are observed in the confined fluid.

\section{Discussion and conclusions}
\label{sec:conc}

In this paper we have presented a simple DFT for describing the bulk and inhomogeneous fluid phase behaviour of a two-dimensional fluid with competing interactions. In Refs.\ \cite{ImperioReattoJPCM2004,ImperioReattoJCP2006} the authors used Monte Carlo computer simulations in the NVT ensemble to study the structure and phase behaviour displayed by the present model fluid. The DFT results presented here are generally in good qualitative agreement with these simulation results, as regards the topology of the phase diagram and the structure of the fluid in the various uniform and inhomogeneous phases. However, the present DFT is not able to describe some properties of the fluid. We discuss these weaknesses now. 

The present theory over estimates by a factor of two the maximum temperature at which microphase-separated phases are expected to occur. This is largely due to the fact that in the present theory we retain for all values of $r$ the form of the potential $w(r)$, given in Eq.\ \eqref{eq:w}, in the mean field contribution to the free energy in Eq.\ \eqref{eq:F}. The value of $w(r)$ for $r<\sigma$ (within the hard core overlap distance) should not contribute to the free energy. That the value of $w(r)$ for $r<\sigma$ does give a contribution to the free energy, is a consequence of the mean field approximation made in constructing the free energy \eqref{eq:F} -- i.e.\ this equation predicts that the free energy of the uniform fluid contains a term proportional to $\alpha=-\int \dr w(r)$, the integral over $w(r)$ for all values of $r$. A better approximation would be to truncate the potential $w(r)$ for $r<\sigma$. This would give much better agreement with the simulation phase diagram \cite{ARCHER19, Archerpreprint}, modifying the value of $\alpha=-\hat{w}(0)$ and effectively just re-scaling the temperature axis, without qualitatively changing any of the results of this paper. We chose not to do this, however, due to that fact that the analysis given in Sec.\ \ref{sec:structure}, for quantities such as $S(k)$ and $k_c$, is much less straight-forward if one truncates $w(r)$, because one then finds that $\hat{w}(k)$ no longer has the simple form given in Eq.\ \eqref{eq:11}.

More substantial differences between the present theoretical predictions and the simulation results are as follows:  In Ref.\ \cite{ImperioReattoJCP2006} the authors show that for temperatures $T<T_{max}$, on increasing the density there is a transition from the vapour phase to a cluster fluid phase -- i.e.\ to a disordered system with no long range ordering of the clusters. On further increasing the density they then found a further transition to a cluster phase with crystalline ordering of the clusters, in line with the predictions from the model presented here. The present theory predicts only a single cluster phase exhibiting crystalline ordering. We believe this is due to the fact our DFT is a simple mean field theory, which neglects certain fluctuation contributions to the free energy \cite{RegueraReissJCP2004}. These neglected contributions must play an important role in the transition from the disordered cluster fluid phase to the ordered cluster phase.

A second difference between results from our DFT and the simulation results in Refs.\ \cite{ImperioReattoJPCM2004,ImperioReattoJCP2006} are that for temperatures $T \lesssim 0.6T_{max}$, the system starts to display crystalline ordering of the particles {\em within} the clusters and stripes. Owing to the fact that we have made a local density approximation in the free energy functional (\ref{eq:F}) for the hard-disk contribution to the free energy, our theory is unable to describe this effect. To describe such freezing effects one must implement a much more sophisticated (non-local) hard-disk reference free energy functional.

\begin{figure}
\includegraphics[width=1.\columnwidth]{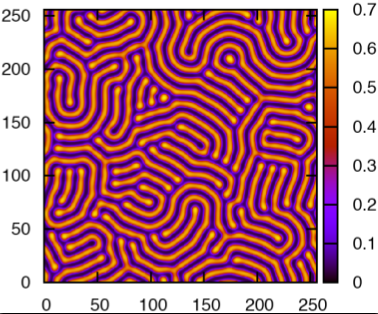}
\includegraphics[width=1.\columnwidth]{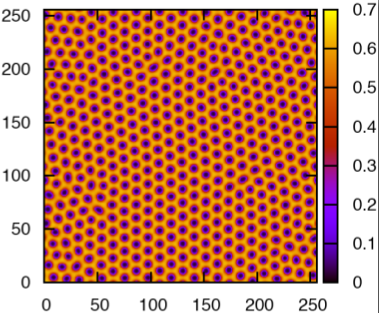}
\caption{\label{fig:defects}
(Color online) Density profiles calculated for $k_BT/\epsilon=0.2$. For the upper profile, the average density is $\rho \sigma^2=0.3$, a state point in the stripe phase. The lower profile is for $\rho \sigma^2=0.45$, a point in the bubble phase. These profiles were obtained by quenching from a homogeneous system with a small amount of random noise added to the density profiles to nucleate the structuring -- see Eq.\ \eqref{eq:rho_noise}. In the stripe phase this results in a density profile containing many defects. Similarly, in the lower density profile this results a number of domains forming in this (periodic) system. The true minimum in the grand potential $\Omega$ corresponds to periodic arrays with no defects.}
\end{figure}

We return now to the issue raised at the end Sec.\ \ref{sec:bulk_phase_behaviour}, regarding the order of the phase transitions. In the present DFT, all the phase transitions between the equilibrium phases are first order with the exception of the transition from the uniform fluid to the stripe phase, which is predicted to be a continuous phase transition. To further understand the nature of the phase transition between the uniform fluid and the stripe phase, we may follow the approach used in Ref.\ \cite{Archerpreprint}. Starting from the DFT \eqref{eq:F}, and assuming that the density profile is of the form
\begin{equation}
\rho(z)=\rho+{\cal  A} \sin(kz),
\end{equation}
the free energy may be expanded in powers of the order parameter ${\cal A}$ in the stripe phase. One finds that the Helmholtz free energy then takes the form
\begin{equation}
{\cal F}[\rho(z)]={\cal F}[\rho]+b_2(\rho,k) {\cal A}^2+b_4(\rho) {\cal A}^4+{\cal O}({\cal A}^6).
\label{eq:F_expand}
\end{equation} 
The coefficient $b_2 \propto a(\eta)+\rho \beta \hat{w}(k)$ [c.f.\ Eq.\ \eqref{eq:S_of_k_model}] and the coefficients $b_n>0$, where $n \geq 4$. The value of the amplitude ${\cal A}$ which minimises the free energy \eqref{eq:F_expand} is the solution to the equation $\partial {\cal F}/\partial {\cal A}=0$. Outside the $\lambda$-line, the coefficient $b_2>0$ for all values of $k$, so that the minimum of the free energy is when ${\cal A}=0$, i.e.\ for the uniform fluid. On the $\lambda$-line itself, $b_2(\rho_\lambda,k_c)=0$ and within the lambda line $b_2(\rho,k_c)<0$, indicating that within the $\lambda$-line the modulated fluid has a free energy that is lower than that of the uniform fluid. At higher temperatures this transition from the uniform fluid to the stripe phase is a second order phase transition, and the transition line is the $\lambda$-line itself. Both the amplitude ${\cal A}$ and the density $\rho$ vary continuously across the transition line (although the first derivative of ${\cal A} (\rho,T)$, does not \cite{Archerpreprint}). On descending to lower temperatures, one finds that the transition between these two phases ceases to be continuous and becomes first order, due to the density and $k$ dependence of the coefficient $b_2(\rho,k)$, and there are two tricritical points connected to one-another by the $\lambda$-line \cite{Archerpreprint}. For temperatures below these tricritical points, both the density $\rho$ and the amplitude ${\cal A}$ vary discontinuously at the phase transition. These tricritical points are not displayed in Fig.\ \ref{fig:phase_diag}, because these points are located in the regions of the phase diagram where the cluster and bubble phases are the equilibrium phases. Thus, in the present two-dimentional system, these tricritical points are not accessible, in contrast, it seems, to the case in three-dimensions \cite{ARCHER21,Archerpreprint}.

In the simulation studies of Imperio and Reatto \cite{ImperioReattoJPCM2004,ImperioReattoJCP2006}, the phase transitions from the uniform fluid to the modulated fluid phases are located by calculating the excess specific heat $C_v^{ex}$ as a function of temperature. This quantity displays a peak and the phase transition is identified with this peak. They find that for low and high densities $\rho$, the peak is well pronounced. However, for intermediate densities around $\rho \sigma^2 \simeq 0.3$, they find that the peak is greatly reduced in amplitude, in comparison with the height of the peak at higher or lower densities \cite{ImperioReattoJCP2006}. Comparing these findings with the present DFT results, we see that where the DFT predicts the phase transition from the uniform fluid to the stripe phase to be second order, coincides with where there is almost no peak in $C_v^{ex}$, and densities where the DFT predicts the transition from the uniform fluid to the modulated fluid to be first order, the simulation results show that there is a pronounced peak in $C_v^{ex}$. In their analysis of the system size scaling of $C_v^{ex}$, Imperio and Reatto conclude that their results for the densities $\rho\sigma^2=0.05$, 0.15 and 0.4 point to the possibility of the phase transition from the modulated fluid phase to the uniform fluid phase being a Kosterlitz-Thouless transition \cite{ImperioReattoJCP2006}. In order to confirm this theoretically, one must go beyond the present mean field DFT theory. One possible route to do this is perhaps by following the approach of Ref.\ \cite{PatsahanCiachJPCM2007}.

In the present two dimensional system, fluctuations are not only important in determining the nature of the phase transitions, but are also important in determining the structure of the bulk modulated phases themselves. Let us consider in particular the stripe phase, although the discussion that follows may also have implications for the cluster and bubble phases. The present DFT predicts that the equilibrium configuration in the stripe phase consists of an array of perfectly parallel stripes, displaying long range order. However, as we have noted already, the present DFT is a mean field theory and neglects certain fluctuation contributions to the free energy. Thus, although the minimum of the free energy corresponds to perfectly parallel stripes, in reality one should find that fluctuations about this minimum will destroy the long range order -- see for example the discussion in Refs.\ \cite{LevinPRL2007, Brazovskii}. In fact, when one seeks for the minimum of the free energy, using as initial guess in the numerical procedure the noisy density profile given in \eqref{eq:rho_noise}, one finds that as the system size $L_x \times L_y$ increases, there are an increasingly large number of metastable minima in the free energy. The density profiles for these minima may contain many defects -- see for example the results displayed in Fig.\ \ref{fig:defects}. These density profiles are calculated for state points in the stripe phase (upper figure) and in the bubble phase (lower figure). Whilst the DFT shows that such configurations corresponds to a minimum in the free energy, the theory does not reveal the height of the free energy barrier between this configuration and any other neighbouring configurations. In order to address the influence of fluctuations and to tackle the question of the size of the barriers between free energy minima, we propose to extend the present theory in the following manner: If one assumes over-damped stochastic equations of motion for the colloidal particles, then following the approach of Ref.\ \cite{Archer} one may develop a dynamical DFT, which predicts the following stochastic equation of motion for the time dependence of the coarse grained fluid density profile:
\begin{equation}
\frac{\partial \rho(\rr,t)}{\partial t} = \nabla\cdot \left[ \rho(\rr,t)\,\nabla \frac{\delta
{\cal F}[\rho]}{\delta \rho(\rr,t)} + \nabla \cdot \sqrt{\rho(\rr,t)}\,\xi(\rr,t)\right],
\label{eq:coarseevol}
\end{equation}
where ${\cal F}[\rho]$ is the free energy functional in Eq.\ \eqref{eq:F} and $\xi(\rr,t)$ is Gaussian random noise field. Such an approach may allow one to determine the influence of fluctuations in the present system and perhaps also to address the question of whether the fluid effectively becomes trapped in configurations such as those displayed in Fig.\ \ref{fig:defects}, as the temperature is decreased -- i.e.\ to determine if and where the glass transition line is in the present system \cite{WestfahletalPRB2001, TarziaConiglioPRE2007}. We plan to pursue this line of investigation in the future.

The sequence of structures displayed by the present model fluid (clusters to stripes to bubbles), has also been observed in the configurations of a subcritical constrained lattice gas (Ising model) \cite{NeuhausHagerJSP2003} and a fluid of rod-like particles \cite{ThompsonPRE2006} -- both of these are two-dimensional systems. Similar behaviour has also been observed in the three-dimensional Lennard-Jones fluid \cite{MacDowelletalJCP2006}. None of these models have long-ranged repulsive interactions between the particles, in contrast to the present system. This sequence of structures is only observed when these systems are constrained within a finite-sized box with periodic boundary conditions and with the number of particles fixed, so that the average density in the box is between that of the coexisting vapour and liquid phases \cite{NeuhausHagerJSP2003, ThompsonPRE2006, MacDowelletalJCP2006}. The cluster/stripe/bubble structures in these systems arise due to the constraint that the number of particles is fixed and the structures are unstable with respect to fluctuations in the density. In contrast to this, in the present system the configurations that we observe are stable equilibrium structures and do survive in an open (grand-cannonical) system. The long range repulsion between the particles stabilises the structures, in just the same way as the long-range repulsion stabilises the structures exhibited by three-dimensional fluids with competing interactions \cite{ARCHER21}. A further difference is that in systems without the long range repulsive interaction between the particles \cite{NeuhausHagerJSP2003, ThompsonPRE2006, MacDowelletalJCP2006}, the size of the clusters/stripes/bubbles depend on the size of the box they are in. However, in the present system, it is the competition between the short range attraction and the long range repulsion that determines the size of the structures and as long as they are within a sufficiently large box, the size of the clusters/stripes/bubbles is independent of the system size.

To conclude, we remind the reader that there are many important applications for systems having colloidal nanoparticles that are confined to a fluid interface, ranging from optical devices to stabilising emulsions \cite{BresmeOettelJPCM2007}. Whilst the present and other studies have gone some way towards understanding the self-assembly of such two-dimensional fluids, there are still several open questions to be addressed \cite{BresmeOettelJPCM2007}.

\acknowledgments

I gratefully acknowledge support from RCUK and to A. Imperio and L. Reatto for several useful discussion.


\begin{thebibliography}{10}
\bibitem{footnote}
We have also checked that the periodic structures do not depend on system size -- i.e.\ one obtains exactly the same results under the transformation $L_x \to 2L_x$.
\end{thebibliography}
\end{document}